\documentclass[
showpacs,preprintnumbers,
nofootinbib,
 amsmath,amssymb,
 aps,
prd,
twocolumn
]{revtex4-1}

\usepackage{graphicx}
\usepackage[margin=1.25in]{geometry}
\usepackage[usenames,dvipsnames]{color}
\usepackage{url}
\usepackage[colorlinks = true,
            linkcolor = blue,
            urlcolor  = blue,
            citecolor = blue,
            anchorcolor = blue]{hyperref}

\begin{document}
\title{Impact of COVID-19 on Astronomy: Two Years In}
\author{Vanessa B\"ohm}
\affiliation{Berkeley Center for Cosmological Physics, University of California, 341 Campbell Hall, Berkeley, CA 94720, USA}
\email{vboehm@berkeley.edu}

\author{Jia Liu}
\affiliation{Kavli IPMU (WPI), UTIAS, The University of Tokyo, Kashiwa, Chiba 277-8583, Japan}
\email{jia.liu@ipmu.jp}
\thanks{equal contribution}

\begin{abstract}
We study the impact of the COVID-19 pandemic on astronomy using public records of astronomical publications.
We show that COVID-19 has had both positive and negative impacts on research in astronomy. We find that the overall output of the field, measured by the yearly paper count, has increased. This is mainly driven by boosted individual productivity seen across most countries, possibly the result of cultural and technological changes in the scientific community during COVID. However, a decreasing number of incoming new researchers is seen in most of the 
countries we studied, indicating larger barriers for new researchers to enter the field or for junior researchers to complete their first project during COVID. Unfortunately, the overall improvement in productivity seen in the field is not equally shared by female astronomers. By fraction, fewer papers are written by women and fewer women are among incoming new researchers in most countries. Even though female astronomers also became more productive during COVID, the level of improvement is smaller than for men. Pre-COVID, female astronomers in the Netherlands, Australia, Switzerland were equally as or even more productive than their male colleagues. During COVID, no single country's female astronomers were able to be more productive than their male colleagues on average.
\end{abstract}

\maketitle

\section{Introduction}

The COVID-19 pandemic has impacted the personal and professional lives of people worldwide. Here, we study the impact of COVID-19 on scientific researches, in particular in the field of astronomy, using public records of astronomical publications from 1950 to February 2022 -- which includes two years during which research worldwide was disrupted by the COVID-19 pandemic. COVID-induced school and work closures, hiring freezes, loss of childcare, increased eldercare duties, and related mental health issues have had undeniably negative impacts on researchers in all fields. However, changes such as increased flexibility in work arrangement, reduced commutes and business trips, as well as improved virtual technologies are potentially favorably for conducting scientific researches. 
Astronomy has well organized and digitalized publication records, making it possible to trace historical trends. Most articles are published as preprints with little delay between the time of production and the time of appearance on the database. In addition, unlike its neighboring fields such as math and physics, which often have alphabetical author ordering, astronomy follows a by-contribution author ranking tradition, making it relatively straightforward to quantify individual authors' contribution and workload. With this targeted study within astronomy, we hope to provide a glimpse into a bigger picture of how COVID-19 may have impacted scientific research. 


In addition, we study the impact of COVID-19 on gender gaps in astronomy. Pre-COVID, large gender gaps already existed in the science, technology, engineering, and mathematics (STEM) fields~\cite{alam_2020} and astronomy saw no exception~\cite{Davenport2014,Reid2014,Patat2016,Caplar2017}. The long-term causes for women to quit STEM jobs---workplace discrimination, lack of encouragement to advance their career, lack of support in balancing work and family, and lack of role models~\cite{funk_parker_2020}---might be amplified during COVID. Earlier work by Ref.~\cite{Myers2020} showed that female scientists and those with young children are disproportionately affected by COVID-19, resulting in 5\% and 17\% larger decline in research time, respectively. 
Despite the fact that institutes have relaxed their rules to help early career scientists cope with COVID-19, Ref.~\cite{NAP26061} reported that many intended gender-neutral responses, such as work-from-home provisions and extensions on evaluations, may in fact exacerbate underlying gender inequalities. 
To study these impacts on women in astronomy, we assign a gender to authors by their given name, and compare the statistics for female astronomers to the general state of the field.


Finally, we study the impact of COVID-19 by country. Policy makers in different countries have made drastically different responses to the pandemic, such as stay-home orders, border restrictions, mask wearing, testing and contact tracing, vaccination, financial support and relief, school and workplace closures~\cite{Hale2021}. We expect these measures to have varied impacts on the local scientific community. 
Within the Italian astronomical community, the submission to arXiv from women is significantly under average for 2020 with respect to the previous years, while that from men is larger by up to 10\% ($3.5\sigma$)~\cite{Inno2020}. Similarly, women, precarious researchers, parents, and expatriates are particularly impacted by COVID-19 in the French astronomical community~\cite{Leboulleux2021}. 
An OECD\footnote{The Organisation for Economic Co-operation and Development: \url{https://www.oecd.org/}.} report compared the policies in 18 member countries and found that countries are in general under-prepared for pandemics~\cite{OECD2022}. It raised the concern of long-term costs and effectiveness of current swift and massive measures such as lockdowns given their impacts on youth and mental health. Overall, they found it too early to assess the true impacts of many of the measures as the crisis is still ongoing. 
Evaluating the best practices for the astronomical society is beyond our expertise. Nevertheless, we take advantage of the rich information in our dataset, in particular the country information embedded in the author affiliations, and apply a by-country analysis, hoping to find some hints to the link between different policies and the well-being of the scientific community. 


This paper is organized as follows. We describe our methodology in Sec.~\ref{sec:method}, including the publication dataset, how we identify individual authors and their gender, with careful anonymization. We then present our results in Sec.~\ref{sec:result}, where we compare the overall output of the field, number of incoming new authors, individual productivity during COVID to pre-COVID, together with related gender disparity measurements. Furthermore, we study the current state for the 25 most populated (w.r.t. number of authors) countries. We conclude in Sec.~\ref{sec:conclude} with the key findings.

\section{Methodology } 
\label{sec:method}

\subsection{Data acquisition}

The dataset used in this study was obtained through the \texttt{SAO/NASA Astrophysics Data System}\footnote{\url{https://ui.adsabs.harvard.edu/}}~(ADS) API in a single download on Feb. 13, 2022. The downloaded data includes all entries on the ADS server of type \textit{article} or \textit{eprint} that were published between Jan. 1, 1950 and Feb. 13, 2022 and included in the \textit{astronomy} database. 

We include both refereed and non-refereed articles (preprints) for the main analysis, since our primary goal is to measure productivity instead of scientific impact, which would be too early to judge in any case for recent publications. 
Including preprints is also necessary to reduce potential biases due to the time-lag between the completion of a paper and the acceptance to a journal, which we expect to be prolonged during COVID-19. In necessary cases, we also apply our analysis to refereed articles only in order to test the robustness of the observed trends. 

We discard publications with more than 16 authors, since these are generally large collaboration papers with alphabetical author ranking. The cutoff at 16 authors is somewhat arbitrary---we would like to include as many papers as possible but exclude large collaboration publications, often with hundreds of authors. The cut at 16 corresponds to keeping about 90\% of the data set. We verified that the presented results are not sensitive to the exact location of this cut.
After this cut, we obtain a total of 1,207,197 publications. For each publication entry, we obtain the title of the article, the publication date, the names of the authors, and their affiliations. 

\subsection{Author identification }

Next, we transform the publication dataset into an author dataset, for all authors that have had at least one first-author publication. In this section, we mainly follow Ref.~\cite{Caplar2017}'s method to identify unique authors. 

While the full family names are usually provided, the given names---including the first and middle names---are often in initials. We separate author entries into three categories: with full given names, with initial-only given names, and without a given name. We first match entries with the same full given + family names. We then match initial-only names to that list, by unique initial + family names combination. For initial-only names without a match in the full given name list, we create a new entry for each unique initial + family name; These entries are not included in the gender analysis, as we are unable to identify the gender based on initials. Finally, we discard any entry without any given name (neither initial nor full), as it is not possible to identify unique authors only by their family name with confidence.

Furthermore, we consider possible changes in family names, often associated with changes in family status such as marriage and divorce. To do so, for entries with compound family names---either joint by hyphen or by space---we search for single family name entries that match either of the individual family names. Once we find a match, we merge the entries if they also have the same given (full or initial-only) names. We also discovered that we are able to further match unique non-Latin names (such as Chinese, Japanese, or Korean names), as some journals allow authors to publish their names in their own language alongside the English versions of their names\footnote{For example: \url{https://authors.aps.org/names.html}.}. We confirm the validity of such procedure by manually comparing other information of the matched entries, such as the paper subject and affiliation. 
One caveat is that a complete change in family name would not be captured in our method. This is unavoidable because it is impossible to identify authors by merely their given names. As the result, such authors will be split into two entries, and a more senior author would be mistaken as two junior authors. 
However, the entries with identified name changes correspond to only 0.4\% of the full database, which suggests that the impact of undiscovered name changes is negligible. 

Of a total of 639,068 author entries,  53\% have full given names (337,449), 47\% have initial-only given names (301,597), and $\ll 1\%$ have no given names (22). We are able to match 40\% of the initial-only entries (120,371) to the full given name entries.  Taking into account family name changes, we found an additional 2,681 matches. In total, we identify 516,304 unique authors, among which 258,889 have at least one first author publication. We use the dataset with the 258,889 first authors for our analysis.



\subsection{Gender identification}


To study the impact of COVID-19 by gender, we identify the gender for the unique author entries (258,889). We use the \texttt{genderize.io} API\footnote{\url{https://genderize.io/}} to assigns a gender and a probability ($P_{\rm gender}$) to an input given name\footnote{We acknowledge the limitation in the tools available to us, where non-binary genders are not identified and hence our genders are limited to \textit{female} and \textit{male} only.}. When an entry has multiple given names, we examine the gender of each name. Usually they return the same gender, in which case we take the highest probability as $P_{\rm gender}$. When the given names return different genders, we take the gender for the first given name, if it has $P_{\rm gender}>80\%$---this is to follow the tradition that the first given name, instead of the middle name, is more often used daily. If the first given name has a low gender identity ($P_{\rm gender}<70\%$)\footnote{For example, gender neutral Latin names such as Jess, Blair, and Taylor, or non-Latin names that can correspond to different genders in the original language, such as Lei in Chinese and Kazumi in Japanese.}, we take the name with the highest $P_{\rm gender}$ and assign that gender and $P_{\rm gender}$ to the entry. 

We are able to identify the gender for 68\% entries (176,102). For the remaining entries with unidentified genders, most are initial-only names (30\%) that are impossible to identify. The rest are rare names without a gender record (2\%). 
We apply a cut of $P_{\rm gender}>80\%$ in our gender-related analysis. 
This leaves us with 30,930 female entries (20\%) and 126,529  male entries (80\%), further reducing our completeness from 68\% to 61\%.




\subsection{Anonymized Author Dataset}
We take great care to anonymize the data as early as possible in the analysis process and remove all identifiable information. After author and gender identification, we remove the author names and replace them with unique author IDs. We do not keep any file that contains author names, but instead only record the gender that was assigned to each IDs. We further replace the authors' affiliations with the countries of the affiliations.

After anonymization, we use the publication dataset to generate an author dataset. The author dataset contains the number of publications in each year between 1950--2022, as well as the author ranking and the country of the affiliation for each publication.  We conduct all our analyses with this condensed and anonymized dataset.


\section{Results}
\label{sec:result}

\subsection{Overall Output in Astronomy}

\begin{figure}
\begin{center}
\includegraphics[width=0.5\textwidth]{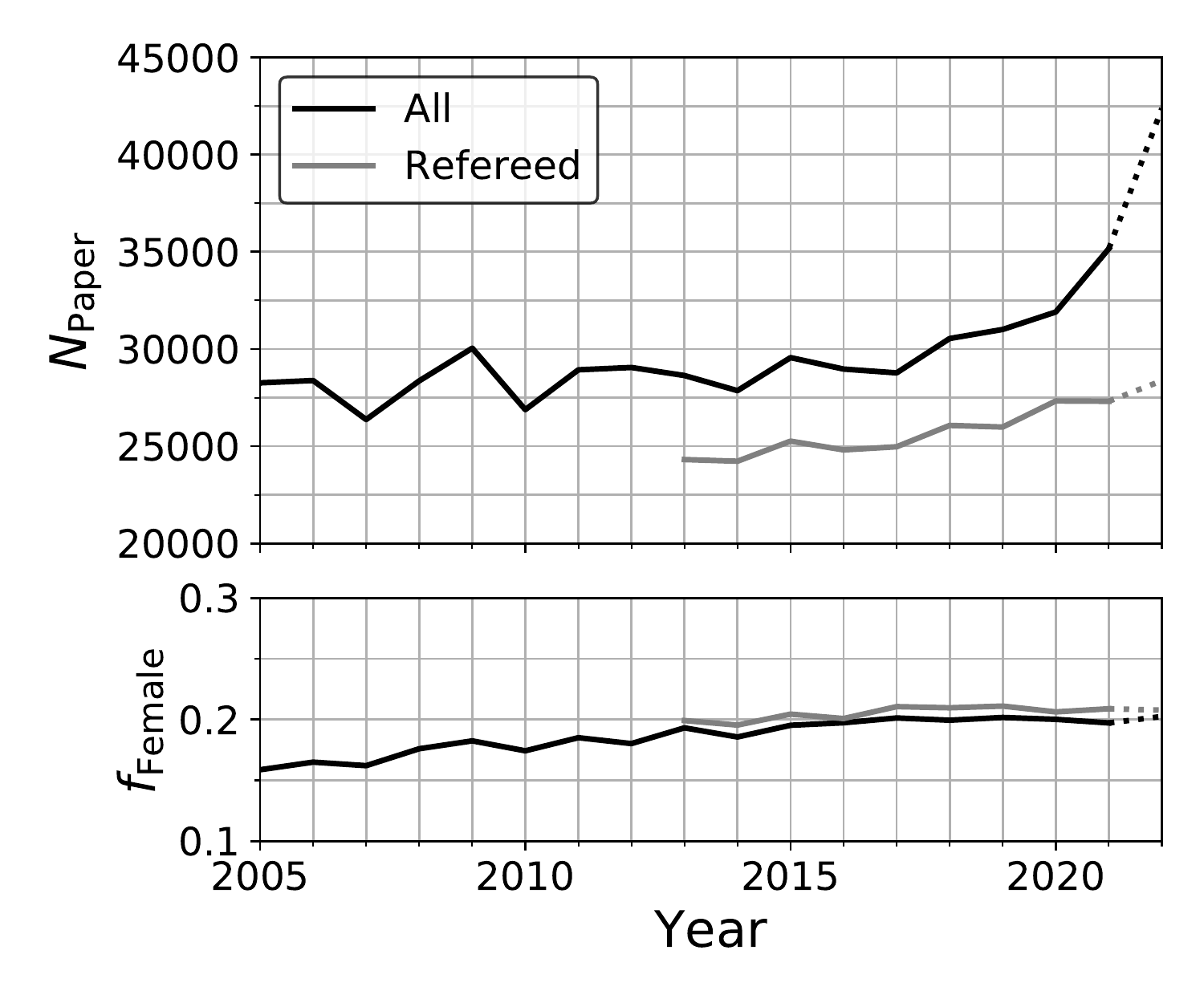}
\end{center}
\caption{\label{fig:Npaper} {\bf Top}: Number of papers in astronomy per year for all papers (black) and refereed papers (grey). {\bf Bottom}: The fraction of papers each year written by female first authors. The dotted lines are projections using partial year data from 2022. The Poisson errors are at the sub-percent level (nearly invisible if shown), and hence the fluctuations seen here are likely due to socioeconomic changes.}
\end{figure}

We show in Fig.~\ref{fig:Npaper} the overall output in the field of astronomy, quantified as the total number of papers per year. The top panel shows the total number of all (refereed+non-refereed) papers per year since 2005. For 2022, we project the full year output by dividing the number of papers published so far by the fraction of the year passed at the time of data download\footnote{We note that publications in astronomy likely have seasonal effects due to job seasons, grant application deadlines, and school calendars. So the simple projection here should be taken with caution.}, shown as dotted lines. The Poisson errors are at the sub-percent level (nearly invisible if shown), and hence the fluctuations seen here are likely due to socioeconomic changes.

We see a general trend of increasing number of papers several years before COVID-19, roughly from 2014. The first 2 years with COVID---2020 and 2021---saw no slowing down of this trend. The projection for 2022 hints on an even faster increase. 
We verify this trend by analyzing only refereed papers, shown as the grey line\footnote{We only include refereed papers for the past 10 years, as the journal review process usually takes only a few months.}. For refereed papers, which is roughly 80\% of all papers, the upward trend remains, though milder. Considering that many authors who submitted their papers to journals did not post their preprints and the likely delayed referee process during COVID, the refereed curve likely reflects the lower limit. Therefore, we conclude that the field of astronomy has been more productive during COVID than pre-COVID.

In the bottom panel of Fig.~\ref{fig:Npaper}, we show the fraction of papers written by female first-authors\footnote{Because we have a large number (30\%) of gender-unidentified authors, we compute the fraction of female author using $f_{\rm Female}=N_{\rm Female}/(N_{\rm Female}+N_{\rm Male})$ instead of $N_{\rm Female}/N_{\rm total}$.}. The fraction of woman-led papers ($\approx$20\%) has not changed significantly in the past 10 years, demonstrating the prolonged gender gap in astronomy. 

The increase in overall output in the field could be due to an increase in new researchers and/or an increase in individual productivity. We investigate the exact sources of the increase below. We will show that except for a few countries that benefited from both, most countries benefit from improved individual productivity alone while incurring a decreasing number of new researchers during COVID. 

\subsection{Incoming New Researchers}

\begin{figure}
\begin{center}
\includegraphics[width=0.5\textwidth]{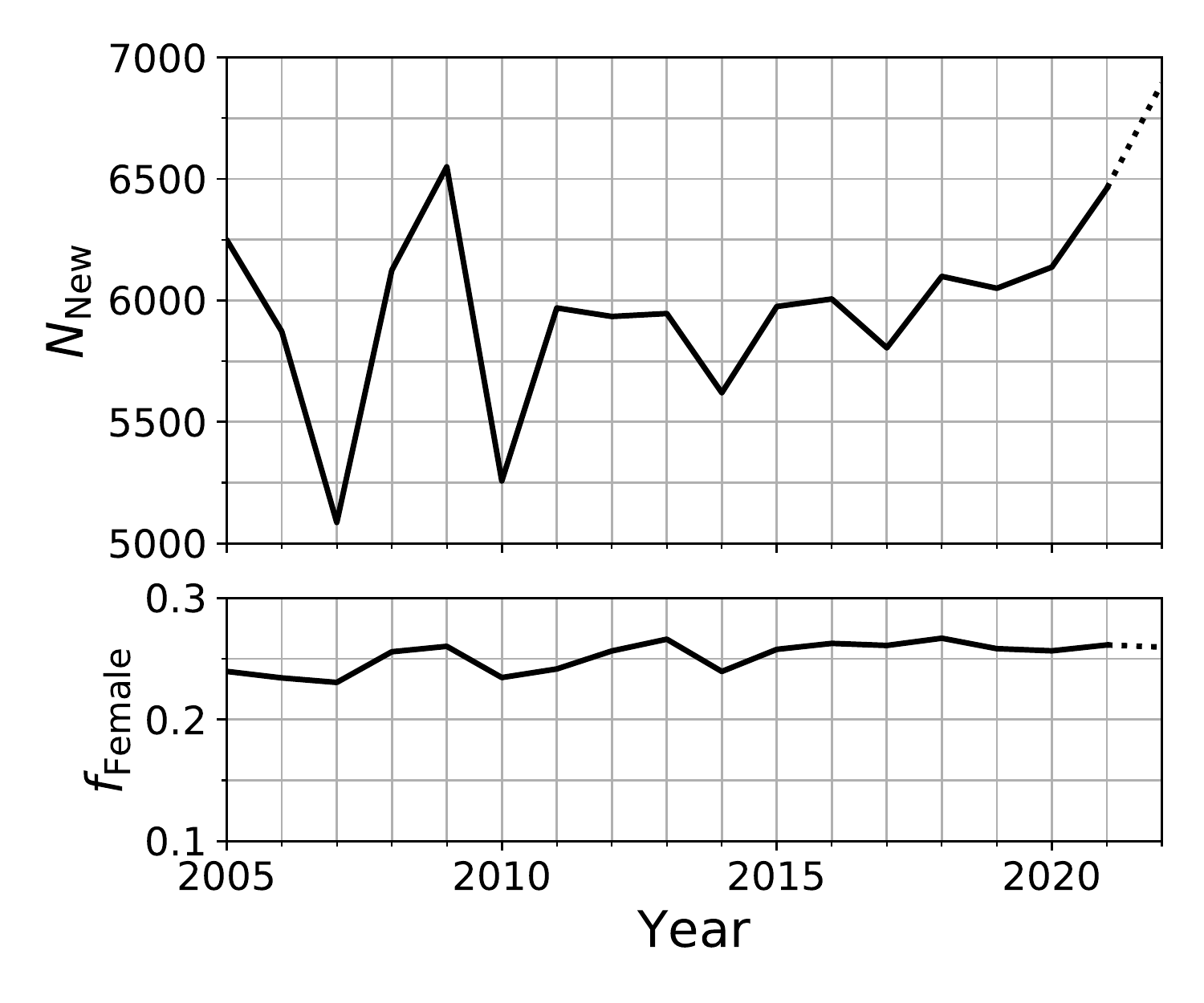}
\end{center}
\caption{\label{fig:Nnew} {\bf Top}: Number of new authors in astronomy per year. {\bf Bottom}: The fraction of women among the new authors. The dotted lines are projections using partial year data from 2022. The Poisson errors are at the percent level (nearly invisible if shown), and hence the fluctuations seen here are likely due to socioeconomic changes.}
\end{figure}

We expect COVID-19 more heavily impact junior researchers in the field. Early career researchers tend to have smaller scientific network, less job securities, and are more likely to have young children. They are hence more vulnerable than more senior, especially tenured, researchers during a crisis. We study the number of incoming new researchers each year. We define new researchers as those who publish their first paper in a given year, disregarding their author ranking. The new authors are likely at early stages of their PhD study, or about to complete their Bachelor's or Master's degree. It is certainly possible that some researchers do not publish their first papers until they become postdocs. Our data do not distinguish the career stages at publication. 

We show the number of new authors per year in the top panel of Fig.~\ref{fig:Nnew}. The Poisson errors are at percent level, and hence any fluctuations seen are likely associated with socioeconomic changes. Compared to before 2020, we see an increase in new authors during COVID. However, as we will show later in Sec.~\ref{sec:country}, this increase is largely driven by only a handful of countries in Asia. The majority of the countries, including the USA and all European countries, see a drop in the number of new authors during COVID. 

We show the fraction of women among all new authors in the bottom panel of Fig.~\ref{fig:Nnew}. We again see no significant improvement for the past 10 years and possibly longer. This indicates strong barriers for women to enter the field of astronomy, likely due to barriers faced earlier during their education. 

\subsection{Individual Productivity}
\label{subsec:ind_prod}

\begin{figure}
\begin{center}
\includegraphics[width=0.5\textwidth]{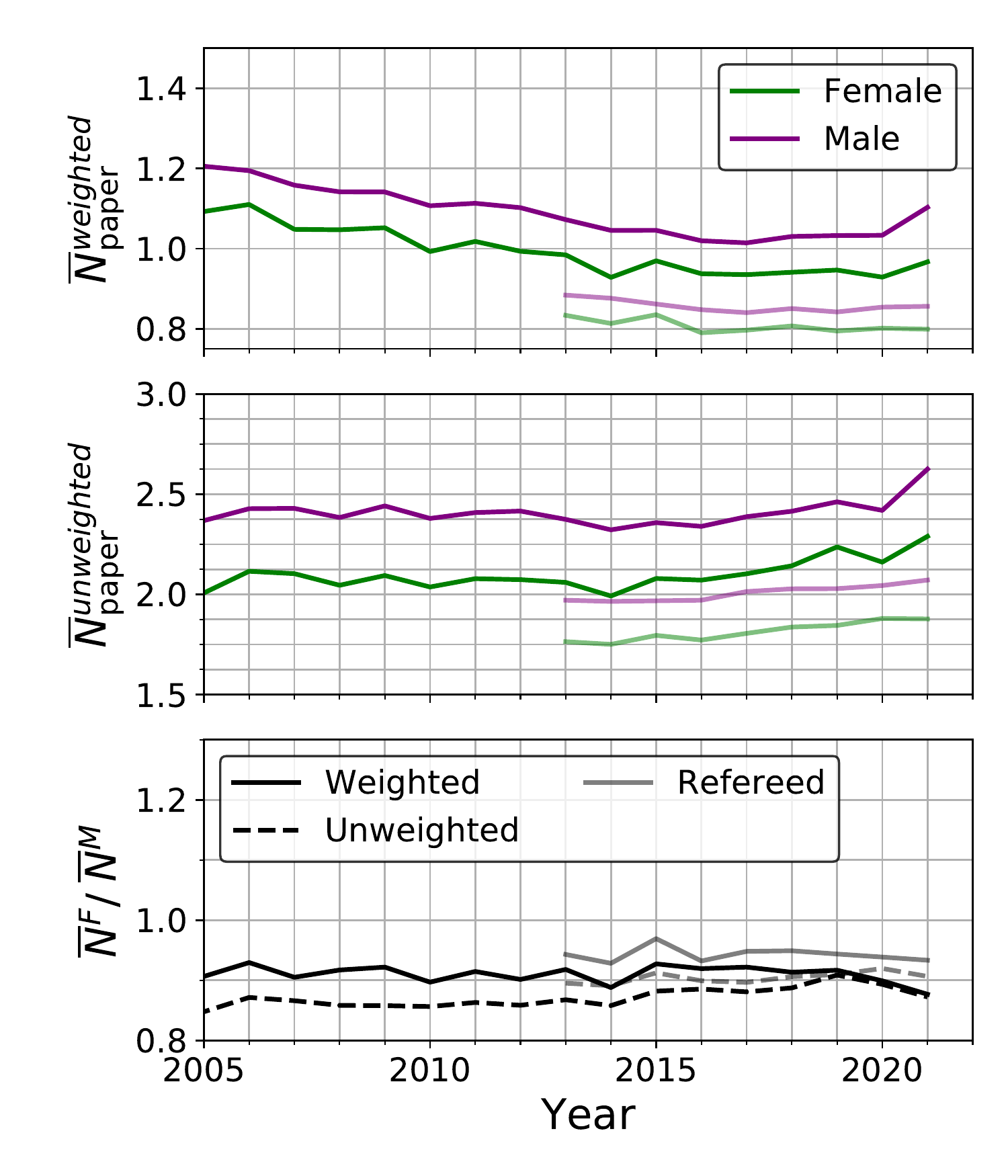}
\end{center}
\caption{\label{fig:Nmean} 
{\bf Top}: Average number of papers 
written by female (green) and male (purple) authors each year, \textit{weighted} by the author rankings. 
{\bf Middle}: \textit{Unweighted} average number of papers written by active authors each year. Active authors are defined as anyone who published in a given year. 
{\bf Bottom}: The ratios between female and male authors in $\overline{N}_{\rm paper}^{\rm weighted}$ (solid) and $\overline{N}_{\rm paper}^{\rm unweighted}$ (dashed). Lighter-colored curves consider only refereed papers.}
\end{figure}

We next study the impact of COVID-19 on the productivity of active researchers on an individual basis. We define ``active'' as anyone who published in a given year. We quantify the level of productivity by counting the number of papers each researcher authored, either weighted by their ranking or unweighted. This simple definition of paper counting does not take into account the quality or impact-level of the paper, as most papers published during COVID have yet to be widely cited. 
We estimate the workload each author has performed that contributes directly to publications using ${N}_{\rm paper}^{\rm weighted}$, 

\begin{equation}
\label{eq:weight}
    {N}_{\rm paper}^{\rm weighted} = \sum_{i} \left(\frac{1}{2}\right)^{(N^i_{\rm rank}-1)}.
\end{equation}
where $N^{i}_{\rm rank}$ is the author ranking and $i$ loops over all papers an author has written in a given year. In other words, we assume none-first authors spend roughly half of the effort of the previous author\footnote{In this scheme, writing 4 third author papers equals to the effort of writing 1 first author paper.}.

In addition, we measure unweighted publication counts ${N}_{\rm paper}^{\rm unweighted}$ for each active researcher. In this scheme, we count all papers equally. Combined with weighted counts from above, unweighted counts provides additional information of collaborative level. A ${N}_{\rm paper}^{\rm unweighted}$ that is significantly larger than ${N}_{\rm paper}^{\rm weighted}$ shows that an author plays supporting roles in multiple collaborative works---likely the case of senior researchers working with many junior students and postdocs. While COVID-19 has cutoff physical contact with collaborators for many, the field has also seen revolutionary changes that made remote collaboration easier than ever. We are interested in the net results of these effects. 

We show in Fig.~\ref{fig:Nmean} both weighted (top) and unweighted (middle) counts for female (green) and male (purple) authors. We also show refereed paper-only in lighter colors. For all papers (solid lines), we see an overall increase in both weighted and unweighted counts during COVID, though an initial drop is visible at the onset of COVID-19 in 2020, except for male authors' ${N}_{\rm paper}^{\rm weighted}$. When considering refereed papers-only, the increase during COVID remains, albeit milder. We conclude that active (female or male) authors' productivity and scientific collaboration have not been reduced by COVID-19. An improvement may even be present, though longer term data will be necessary to confirm it.

In the bottom panel of Fig.~\ref{fig:Nmean}, we show the gender gap in productivity. Female authors are producing $\approx 10\%$ less papers than male authors at all time (black solid lines). This persisting trend hints at social and systemic discrimination impacting the scientific productivity of women~\cite{alam_2020}. 
In Sec.~\ref{sec:country}, we will show that female researchers in some countries consistently published more than or at similar rates as male researchers pre-COVID, demonstrating that female researchers are indeed competent when provided with adequate support. Furthermore, the gender disparity is worse when taken collaborative work into consideration (dashed lines). This trend could be due to the fact that female authors are at earlier career stages on average, and still need to establish their scientific network, but it could also point towards barriers to networking and joining collaborative research. The trend remains in the refereed paper-only analysis (lighter color), though with a slightly milder gender disparity. Female astronomers' productivity (weighted paper counts) dropped from 92\% pre-COVID to 89\% during COVID, compared to that of male astronomers. The drop for unweighted paper counts is smaller, from 88\% pre-COVID to 89\% during COVID. The drop is less obvious for refereed papers, possibly due to delayed referee process during COVID.

\subsection{Increase in Productivity by Career Stage and Gender}
We further study the interesting phenomenon of increase in productivity during COVID, by separating authors by their career stage. We compare the average productivity of researchers during COVID (defined as the time interval from February 1, 2020 to January 31, 2022) to the average productivity during similar time windows before COVID.

We consider a total of 5 $\times$ two-year time intervals from 2010 to 2019, which we treat as control experiments for comparison with the pandemic time interval during the years 2020 and 2021. For each of these time intervals, we measure the average productivity of active male and female researchers at different career stages. We define \textit{active} as having published either during or in the last two years before the time interval under consideration. A detailed description of the data processing and figures of the monthly productivity in time intervals before and during COVID are provided in Appendix~\ref{app:A}. 

Here we summarize the result by means of the excess productivity, $EP$, where $EP$ is defined as the difference between the productivity during COVID and the average productivity in similar time intervals pre-COVID,
\begin{equation}
    EP = \overline{N}^{\rm covid}_{\rm paper} - \frac{1}{N_{\mathrm{intervals}}} \sum_{y \in \mathrm{pre-covid}} \overline{N}^{y}_{\rm paper}.
\end{equation}
Standard errors are computed by jackknifing on the pre-COVID data. We report the $EP$ divided by the estimated standard error $\sigma_{EP}$, for different career stages and genders. We further compute the ratio of the excess productivity between male and female researchers. Results for first author publications are listed in Table~\ref{tab:monthly_1} and for all publication weighted by author ranking in Table~\ref{tab:monthly_all}. 
\begin{table}[!ht]
    \centering
    \begin{tabular}{|l|c|c|c|c|c|}
    \hline
        career stage & $\frac{EP}{\sigma_{EP}}$ Male & $\frac{EP}{\sigma_{EP}}$ Female & $\frac{EP_{\mathrm{Male}}}{EP_{\mathrm{Female}}}$  \\ \hline
        1-6 yrs & 2.73 & 1.74 & 1.72$\pm$1.17\\ \hline
        6-11 yrs & 1.63 & 1.54 & 1.55$\pm$1.39\\ \hline
        11-16 yrs & 1.74 & 0.49 & 3.46$\pm$7.30 \\ \hline
        16-21 yrs & 0.54 & -0.37 & -0.98$\pm$3.25\\ \hline
    \end{tabular}
\caption{\label{tab:monthly_1}Excess productivity, $EP$, for first author publications during COVID as compared to the average productivity in the years 2010-2019. We report results separated by gender and career stage.}
\end{table}
\begin{table}[!ht]
    \centering
    \begin{tabular}{|l|c|c|c|c|c|}
    \hline
        career stage & $\frac{EP}{\sigma_{EP}}$ Male & $\frac{EP}{\sigma_{EP}}$ Female & $\frac{EP_{\mathrm{Male}}}{EP_{\mathrm{Female}}}$  \\ \hline
        1-6 yrs & 2.16 & 1.55 & 1.51$\pm$1.20\\ \hline
        6-11 yrs & 1.66 & 1.56 & 1.50$\pm$1.32\\ \hline
        11-16 yrs & 3.17 & 1.76 & 1.56$\pm$1.01 \\ \hline
        16-21 yrs & 1.69 & 0.81 & 1.37$\pm$1.86\\ \hline
    \end{tabular}
\caption{\label{tab:monthly_all}Excess productivity, $EP$, for all publications during COVID weighted by author ranking (eq.~\ref{eq:weight}) compared to the average productivity in the years 2010-2019. We report results separated by gender and career stage.}
\end{table}
We find a general increase in productivity. However, this increase is not equally shared among different genders and career stages. Earlier career stages seem to have seen the greatest increase in productivity, but also exhibit greater (more significant) disparities between men and women. An explanation for this finding could be unequal shares in child and elderly care responsibilities.

We use both, refereed and non-refereed, publications in this analysis, since we expect a several month delay between production and completion of the referee process. The observed trend is still the same for refereed articles, but less significant (Appendix~\ref{app:A}).

\subsection{Idling Active Authors}

\begin{figure}
\begin{center}
\includegraphics[width=0.5\textwidth]{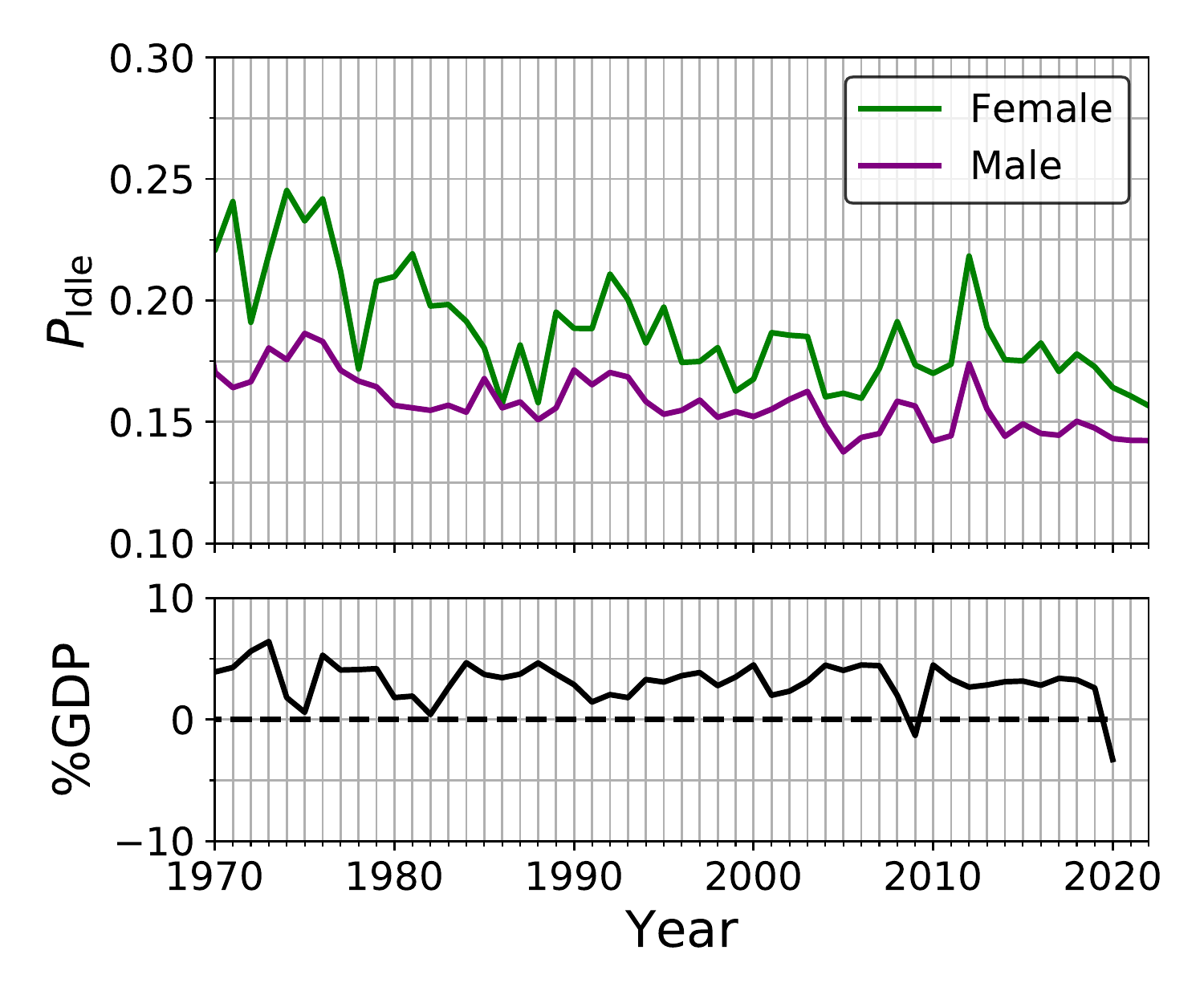}
\end{center}
\caption{\label{fig:idle} {\bf Top}: The fraction of idle authors (defined as no paper in the past 2 years) among the previously active authors (defined as producing $N_{\rm paper}^{\rm weighted}>1$ in the previous 2 years). Active female (green) authors are more likely to idle than active male (purple) authors at all time. {\bf Bottom}: The world GDP growth is shown~\cite{WB2021}, to illustrate the possible anti-correlation of $P_{\rm idle}$ with socioeconomic changes. 
The COVID-19 induced recession is the only one (among 6 notable recessions) since 1970 without an associated peak in $P_{\rm idle}$ for either gender.}
\end{figure}


Next, we turn our focus to the most vulnerable groups, who are likely already adversely affected by COVID-19---the previously active authors who became idle during COVID. We define this population to be those who were productive during the 2 years immediately before COVID with ${N}_{\rm paper}^{\rm weighted}>1$, but have not published any paper for the past 2 years (or during COVID). 
For comparison, we apply the analysis to as early as 1970, using data from 4 years before a specific year, in order to find historical trends potentially related to previous global crisis.

We show the probability for an active researcher to become idle in Fig.~\ref{fig:idle}, for both female (green) and male (purple) authors. We see that female authors have a much higher chance to turn idle (often it means quitting the field) than male authors, at all time. To investigate the potential correlation with world economy, we show the world Gross Domestic Product (GDP) growth in the bottom panel~\cite{WB2021}. We find almost one-on-one correspondences between past recessions (e.g. 1975, 1980--1982, 1991-1993, 2001, 2008--2009) and peaks in $P_{\rm Idle}$ for female authors. The trend is similar (though milder) for male authors. The recent (2008--2009) global financial crisis may have two associated peaks (2008 and 2012).

Surprisingly, we do not see an increase in author idling during the current COVID-induced recession, in contrary to all previous recessions in the past 50 years. 
This could be due to the limited time period of COVID data and that a future peak is yet to be seen. However, judging from how previous $P_{\rm Idle}$ peaks are closely associated with corresponding crisis (on or even 1 year before the year when GDP drops to local minimum) and that we even see a mild drop in idling rate 2 years during COVID, we suspect the impact of COVID-19 may be beyond the usual economic crisis. 
Changes in the cultural and social aspects of astronomical research, e.g. reduced commutes and business trips and increased flexibility in work modes may have allowed otherwise busy researchers to focus on research and become more productive during COVID. 


\subsection{By Country}
\label{sec:country}
\begin{figure*}
\begin{center}
\includegraphics[width=0.98\textwidth]{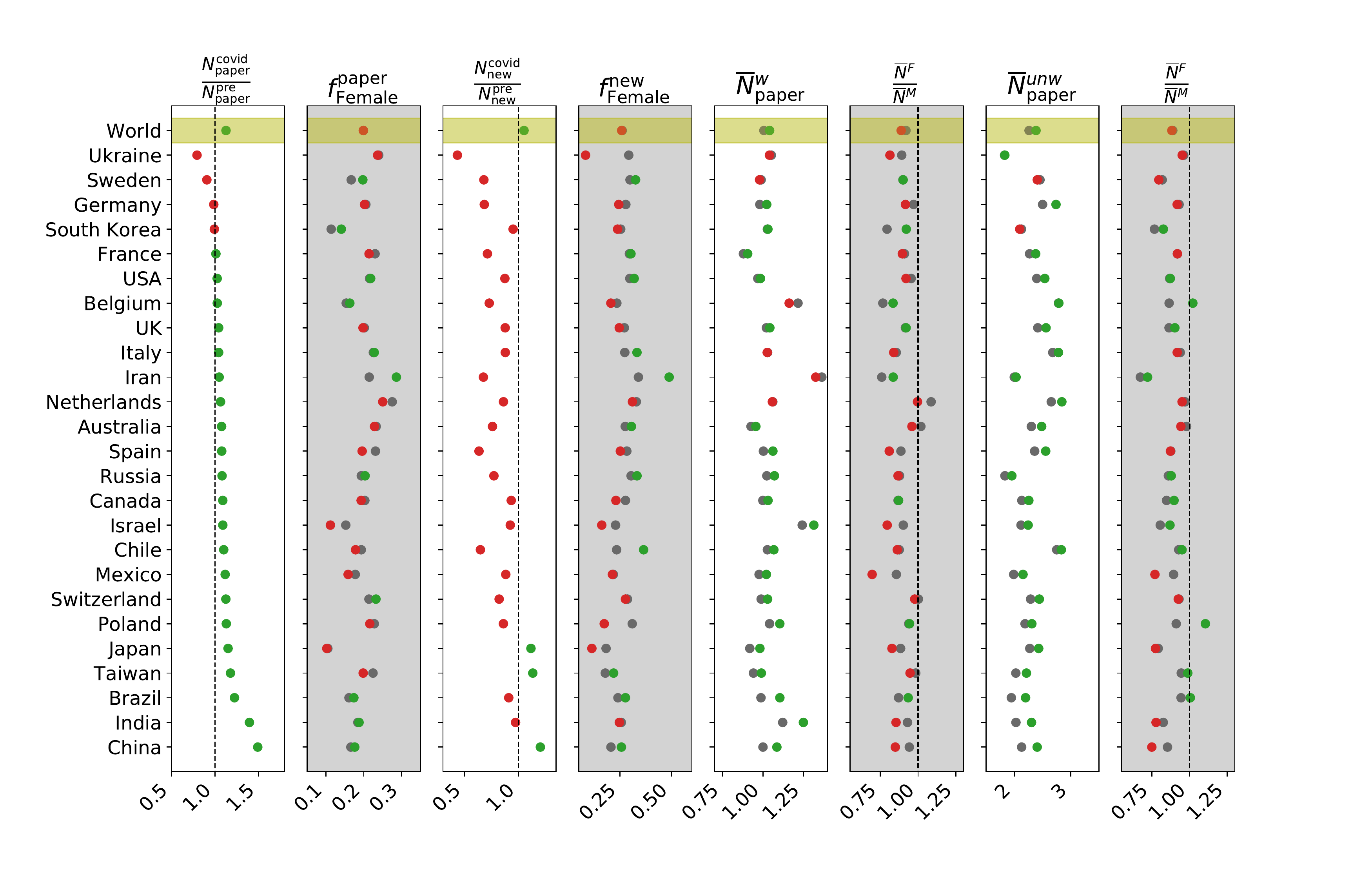}
\end{center}
\caption{\label{fig:country} Impact of COVID-19 on astronomy by country measured in 4 statistics (white columns), each paired with a corresponding gender-disparity measurement (the adjacent grey column to the right). {\bf Grey} points are pre-COVID and  {\bf Red} ({\bf green}) points are during COVID that are worse (better) than pre-COVID. The World statistics are shown in the first row. 
{\bf Cols. 1 \& 2 (overall output)}: Ratio of the number of papers per year during COVID to pre-COVID, and corresponding fraction of female authors.
{\bf Cols. 3 \& 4 (new researchers)}: Ratio of the number of new authors per year during COVID to pre-COVID, and corresponding fraction of new female authors.
{\bf Cols. 5 \& 6 (individual productivity)}: Weighted average number of papers per active author $\overline{N}_{\rm paper}^{\rm weighted}$, where none-first author papers are down-weighted, and the corresponding ratio of that between female to male authors.
{\bf Cols. 7 \& 8 (individual productivity and network size)}: Unweighted average number of papers per active author $\overline{N}_{\rm paper}^{\rm unweighted}$, where the author ranking is ignored in counting papers, and the corresponding ratio of that between female to male authors. We only consider countries with more than 1,000 authors (25 in total). The error bars are not shown for legibility. For guidance, the Poisson errors are between 1\% level for the countries with most authors to 10\% level for the ones with fewer authors.
}
\end{figure*}

Finally, we study the same statistics for authors in different countries. In response to the pandemic, governments in different countries implemented drastically different policies. 
Analysis by country could reflect the impact of different COVID responses on the scientific community. We use the country of the most recent affiliation for each author. 
There are in total 141 countries in our database. We only include countries with more than 1000 identified authors, to ensure that Poisson noise is at most at the 10\% level, resulting in total 25 countries. The selected countries cover a wide range in geography, culture, economic development, social welfare, scientific priority, and COVID-19 policies. 
We compute the same statistics as above (for World) pre- and during COVID for each country. For pre-COVID statistics, we use the average of 5 year data immediately before COVID (2015--2019). For during COVID statistics, we use the average of 2 year data during COVID (2020 and 2021). We do not project for 2022 using partial year data, due to large noises in country-dependent seasonal effects. 

Fig.~\ref{fig:country} shows the four statistics---overall output, number of new authors, individual productivity, and individual productivity+network---in ratios of post- to pre-COVID. Each statistic is shown in pairs of white and grey columns, for the general population and the corresponding gender disparity measurement, respectively. We rank order the countries by the first column (overall output) and show the World's statistics in the first row (shaded in yellow) to guide visual comparison. Grey points are for pre-COVID measurements, while red (green) points show during-COVID values that are worse (better) than pre-COVID. 

First, from the first column, which shows the ratios of total number of papers per year post- to pre-COVID, majority of the countries see an increase in scientific output during COVID (except for a few countries, notably Ukraine and Sweden). To find the drive of this increase, we show in column 3 that most countries actually see a decreasing number of new authors than pre-COVID, and hence incoming new authors are not responsible for the increased output worldwide. Rather, we see an improved individual productivity for most countries (columns 5 \& 7), driving up the overall world paper output. 

Despite the overall increase of the scientific outputs, more than half of the countries see a worsened situation for women. Column 2 shows the fraction of papers from female first authors pre- and during COVID. In general, only 20\% papers are written by female first authors.  During COVID, 14 out of 25 countries saw an even smaller fraction. The exact reasons for this trend are not deducible from our analysis --- declines in the scientific output of female researchers are happening in both better-than-average countries (e.g. the Netherlands) and worse-than-average countries (e.g. Japan and Isreal). 

Next, when examining the number of incoming new authors (column 3), we found that the increasing world average is mostly driven by only a handful countries, namely China, Japan, and Taiwan\footnote{We hypothesise the cause may be that these countries applied rapid responses to COVID-19 early on, ranging from border restrictions, close contract tracing, face covering requirements, to lockdowns. These are also the regions suffered most severely during the previous 2002--2004 SARS outbreak and hence they might be more experienced with health crisis.}. 
Most other countries see a reduced number of new authors, dropping down to as low as 50--70\% of pre-COVID level. When examining the fraction of women among new authors, we see that they roughly stay at the same value as pre-COVID (column 4), indicating no additional barriers for women to enter astronomy were present during COVID.

Finally, when studying the individual productivity, we find that most countries see an improvement in individual productivity (column 5) and collaboration (column 7). However, this improvement is not equally shared by female researchers. Individual productivity has declined for women (when compared to men) in 17 out of the 25 countries. More strikingly, no single country's female researchers are able to be as productive as men during COVID, including the previously 110\% out-performing female astronomers in the Netherlands. A similar situation is seen when taking into consideration scientific networking (column 8) where many countries experienced a decline for women. These findings indicate that COVID-19 has taken a higher toll on female researchers. 

To summarize our by-country analysis, we see an overall increased output driven by existing researcher's  productivity, rather than an influx of new authors. However, the disadvantages female researches faced before the pandemic have remained or even worsened during COVID. 


\section{Conclusion}
\label{sec:conclude}

We study the impact of COVID-19 on astronomy by measuring (1) the overall output of the field, (2) incoming new researchers, (3) individual productivity and network, and (4) active researcher's likelihood to idle. We examine the corresponding gender gaps. We also apply by-country analysis for 25 countries that have more than 1000 authors. Our key findings are:
\begin{itemize}
    \item The overall output of the field, represented by the number of publications, has increased during COVID, for the whole world as well as for most countries (21 out of 25).  
    \item Most countries (22 out of 25) see a decreasing number of incoming new researchers, except for Japan, Taiwan, and China, indicating larger barriers during COVID for new researchers to enter the field or for junior researchers to complete their first project. 
    \item Most countries see boosted individual productivity, both in terms of increased contribution to scientific papers and larger collaboration network, hinting on the positive impacts associated with cultural and technological changes induced by the pandemic. 
    \item While the world sees an improvement in researchers' productivity, the gender disparity has been widened---smaller fraction of papers are written by women, and women make up a smaller fraction of incoming new researchers. Even though women are also more productive during COVID, the level of improvement is smaller than for men. Pre-COVID, female astronomers in the Netherlands, Australia, Switzerland were equally as or more productive than their male colleagues. During COVID, no single country’s female astronomers are able to be more productive than their male colleagues on average.

\end{itemize}

\begin{acknowledgments}
We thank Maria Charisi, Koun Choi, Zoltan Haiman, Kohei Inayoshi, Yukari Ito, Inna Okounkova, Lilan Yang for helpful discussions and comments.
\end{acknowledgments}


\bibliography{ref}

\begin{thebibliography}{13}%
\makeatletter
\providecommand \@ifxundefined [1]{%
 \@ifx{#1\undefined}
}%
\providecommand \@ifnum [1]{%
 \ifnum #1\expandafter \@firstoftwo
 \else \expandafter \@secondoftwo
 \fi
}%
\providecommand \@ifx [1]{%
 \ifx #1\expandafter \@firstoftwo
 \else \expandafter \@secondoftwo
 \fi
}%
\providecommand \natexlab [1]{#1}%
\providecommand \enquote  [1]{``#1''}%
\providecommand \bibnamefont  [1]{#1}%
\providecommand \bibfnamefont [1]{#1}%
\providecommand \citenamefont [1]{#1}%
\providecommand \href@noop [0]{\@secondoftwo}%
\providecommand \href [0]{\begingroup \@sanitize@url \@href}%
\providecommand \@href[1]{\@@startlink{#1}\@@href}%
\providecommand \@@href[1]{\endgroup#1\@@endlink}%
\providecommand \@sanitize@url [0]{\catcode `\\12\catcode `\$12\catcode
  `\&12\catcode `\#12\catcode `\^12\catcode `\_12\catcode `\%12\relax}%
\providecommand \@@startlink[1]{}%
\providecommand \@@endlink[0]{}%
\providecommand \url  [0]{\begingroup\@sanitize@url \@url }%
\providecommand \@url [1]{\endgroup\@href {#1}{\urlprefix }}%
\providecommand \urlprefix  [0]{URL }%
\providecommand \Eprint [0]{\href }%
\providecommand \doibase [0]{http://dx.doi.org/}%
\providecommand \selectlanguage [0]{\@gobble}%
\providecommand \bibinfo  [0]{\@secondoftwo}%
\providecommand \bibfield  [0]{\@secondoftwo}%
\providecommand \translation [1]{[#1]}%
\providecommand \BibitemOpen [0]{}%
\providecommand \bibitemStop [0]{}%
\providecommand \bibitemNoStop [0]{.\EOS\space}%
\providecommand \EOS [0]{\spacefactor3000\relax}%
\providecommand \BibitemShut  [1]{\csname bibitem#1\endcsname}%
\let\auto@bib@innerbib\@empty
\bibitem [{\citenamefont {Alam}(2020)}]{alam_2020}%
  \BibitemOpen
  \bibfield  {author} {\bibinfo {author} {\bibfnamefont {A.}~\bibnamefont
  {Alam}},\ }\href
  {https://www.unicef.org/globalinsight/stories/mapping-gender-equality-stem-school-work}
  {\enquote {\bibinfo {title} {Mapping gender equality in stem from school to
  work},}\ } (\bibinfo {year} {2020})\BibitemShut {NoStop}%
\bibitem [{\citenamefont {{Davenport}}\ \emph {et~al.}(2014)\citenamefont
  {{Davenport}}, \citenamefont {{Fouesneau}}, \citenamefont {{Grand}},
  \citenamefont {{Hagen}}, \citenamefont {{Poppenhaeger}},\ and\ \citenamefont
  {{Watkins}}}]{Davenport2014}%
  \BibitemOpen
  \bibfield  {author} {\bibinfo {author} {\bibfnamefont {J.~R.~A.}\
  \bibnamefont {{Davenport}}}, \bibinfo {author} {\bibfnamefont
  {M.}~\bibnamefont {{Fouesneau}}}, \bibinfo {author} {\bibfnamefont
  {E.}~\bibnamefont {{Grand}}}, \bibinfo {author} {\bibfnamefont
  {A.}~\bibnamefont {{Hagen}}}, \bibinfo {author} {\bibfnamefont
  {K.}~\bibnamefont {{Poppenhaeger}}}, \ and\ \bibinfo {author} {\bibfnamefont
  {L.~L.}\ \bibnamefont {{Watkins}}},\ }\href@noop {} {\bibfield  {journal}
  {\bibinfo  {journal} {arXiv e-prints}\ ,\ \bibinfo {eid} {arXiv:1403.3091}}
  (\bibinfo {year} {2014})},\ \Eprint {http://arxiv.org/abs/1403.3091}
  {arXiv:1403.3091 [physics.soc-ph]} \BibitemShut {NoStop}%
\bibitem [{\citenamefont {{Reid}}(2014)}]{Reid2014}%
  \BibitemOpen
  \bibfield  {author} {\bibinfo {author} {\bibfnamefont {I.~N.}\ \bibnamefont
  {{Reid}}},\ }\href {\doibase 10.1086/678964} {\bibfield  {journal} {\bibinfo
  {journal} {The Astronomical Society of the Pacific}\ }\textbf {\bibinfo
  {volume} {126}},\ \bibinfo {pages} {923} (\bibinfo {year} {2014})},\ \Eprint
  {http://arxiv.org/abs/1409.3528} {arXiv:1409.3528 [astro-ph.IM]} \BibitemShut
  {NoStop}%
\bibitem [{\citenamefont {{Patat}}(2016)}]{Patat2016}%
  \BibitemOpen
  \bibfield  {author} {\bibinfo {author} {\bibfnamefont {F.}~\bibnamefont
  {{Patat}}},\ }\href@noop {} {\bibfield  {journal} {\bibinfo  {journal} {The
  Messenger}\ }\textbf {\bibinfo {volume} {165}},\ \bibinfo {pages} {2}
  (\bibinfo {year} {2016})},\ \Eprint {http://arxiv.org/abs/1610.00920}
  {arXiv:1610.00920 [physics.soc-ph]} \BibitemShut {NoStop}%
\bibitem [{\citenamefont {Caplar}\ \emph {et~al.}(2017)\citenamefont {Caplar},
  \citenamefont {Tacchella},\ and\ \citenamefont {Birrer}}]{Caplar2017}%
  \BibitemOpen
  \bibfield  {author} {\bibinfo {author} {\bibfnamefont {N.}~\bibnamefont
  {Caplar}}, \bibinfo {author} {\bibfnamefont {S.}~\bibnamefont {Tacchella}}, \
  and\ \bibinfo {author} {\bibfnamefont {S.}~\bibnamefont {Birrer}},\ }\href
  {\doibase 10.1038/s41550-017-0141} {\bibfield  {journal} {\bibinfo  {journal}
  {Nature Astronomy}\ }\textbf {\bibinfo {volume} {1}} (\bibinfo {year}
  {2017}),\ 10.1038/s41550-017-0141}\BibitemShut {NoStop}%
\bibitem [{\citenamefont {Funk}\ and\ \citenamefont
  {Parker}(2020)}]{funk_parker_2020}%
  \BibitemOpen
  \bibfield  {author} {\bibinfo {author} {\bibfnamefont {C.}~\bibnamefont
  {Funk}}\ and\ \bibinfo {author} {\bibfnamefont {K.}~\bibnamefont {Parker}},\
  }\href
  {https://www.pewresearch.org/social-trends/2018/01/09/women-and-men-in-stem-often-at-odds-over-workplace-equity/}
  {\enquote {\bibinfo {title} {Women and men in stem often at odds over
  workplace equity},}\ } (\bibinfo {year} {2020})\BibitemShut {NoStop}%
\bibitem [{\citenamefont {Myers}\ \emph {et~al.}(2020)\citenamefont {Myers},
  \citenamefont {Tham}, \citenamefont {Yin}, \citenamefont {Cohodes},
  \citenamefont {Thursby}, \citenamefont {Thursby}, \citenamefont {Schiffer},
  \citenamefont {Walsh}, \citenamefont {Lakhani},\ and\ \citenamefont
  {Wang}}]{Myers2020}%
  \BibitemOpen
  \bibfield  {author} {\bibinfo {author} {\bibfnamefont {K.~R.}\ \bibnamefont
  {Myers}}, \bibinfo {author} {\bibfnamefont {W.~Y.}\ \bibnamefont {Tham}},
  \bibinfo {author} {\bibfnamefont {Y.}~\bibnamefont {Yin}}, \bibinfo {author}
  {\bibfnamefont {N.}~\bibnamefont {Cohodes}}, \bibinfo {author} {\bibfnamefont
  {J.~G.}\ \bibnamefont {Thursby}}, \bibinfo {author} {\bibfnamefont {M.~C.}\
  \bibnamefont {Thursby}}, \bibinfo {author} {\bibfnamefont {P.}~\bibnamefont
  {Schiffer}}, \bibinfo {author} {\bibfnamefont {J.~T.}\ \bibnamefont {Walsh}},
  \bibinfo {author} {\bibfnamefont {K.~R.}\ \bibnamefont {Lakhani}}, \ and\
  \bibinfo {author} {\bibfnamefont {D.}~\bibnamefont {Wang}},\ }\href {\doibase
  10.1038/s41562-020-0921-y} {\bibfield  {journal} {\bibinfo  {journal} {Nature
  Human Behaviour}\ }\textbf {\bibinfo {volume} {4}},\ \bibinfo {pages} {880}
  (\bibinfo {year} {2020})}\BibitemShut {NoStop}%
\bibitem [{\citenamefont {{National Academies of Sciences, Engineering, and
  Medicine}}(2021)}]{NAP26061}%
  \BibitemOpen
  \bibfield  {author} {\bibinfo {author} {\bibnamefont {{National Academies of
  Sciences, Engineering, and Medicine}}},\ }\href {\doibase 10.17226/26061}
  {\emph {\bibinfo {title} {The Impact of COVID-19 on the Careers of Women in
  Academic Sciences, Engineering, and Medicine}}},\ edited by\ \bibinfo
  {editor} {\bibfnamefont {E.}~\bibnamefont {Higginbotham}}\ and\ \bibinfo
  {editor} {\bibfnamefont {M.~L.}\ \bibnamefont {Dahlberg}}\ (\bibinfo
  {publisher} {The National Academies Press},\ \bibinfo {address} {Washington,
  DC},\ \bibinfo {year} {2021})\BibitemShut {NoStop}%
\bibitem [{\citenamefont {Hale}\ \emph {et~al.}(2021)\citenamefont {Hale},
  \citenamefont {Angrist}, \citenamefont {Goldszmidt}, \citenamefont {Kira},
  \citenamefont {Petherick}, \citenamefont {Phillips}, \citenamefont {Webster},
  \citenamefont {Cameron-Blake}, \citenamefont {Hallas}, \citenamefont
  {Majumdar},\ and\ \citenamefont {Tatlow}}]{Hale2021}%
  \BibitemOpen
  \bibfield  {author} {\bibinfo {author} {\bibfnamefont {T.}~\bibnamefont
  {Hale}}, \bibinfo {author} {\bibfnamefont {N.}~\bibnamefont {Angrist}},
  \bibinfo {author} {\bibfnamefont {R.}~\bibnamefont {Goldszmidt}}, \bibinfo
  {author} {\bibfnamefont {B.}~\bibnamefont {Kira}}, \bibinfo {author}
  {\bibfnamefont {A.}~\bibnamefont {Petherick}}, \bibinfo {author}
  {\bibfnamefont {T.}~\bibnamefont {Phillips}}, \bibinfo {author}
  {\bibfnamefont {S.}~\bibnamefont {Webster}}, \bibinfo {author} {\bibfnamefont
  {E.}~\bibnamefont {Cameron-Blake}}, \bibinfo {author} {\bibfnamefont
  {L.}~\bibnamefont {Hallas}}, \bibinfo {author} {\bibfnamefont
  {S.}~\bibnamefont {Majumdar}}, \ and\ \bibinfo {author} {\bibfnamefont
  {H.}~\bibnamefont {Tatlow}},\ }\href {\doibase 10.1038/s41562-021-01079-8}
  {\bibfield  {journal} {\bibinfo  {journal} {Nature Human Behaviour}\ }\textbf
  {\bibinfo {volume} {5}},\ \bibinfo {pages} {529} (\bibinfo {year}
  {2021})}\BibitemShut {NoStop}%
\bibitem [{\citenamefont {Inno}\ \emph {et~al.}(2020)\citenamefont {Inno},
  \citenamefont {Rotundi},\ and\ \citenamefont {Piccialli}}]{Inno2020}%
  \BibitemOpen
  \bibfield  {author} {\bibinfo {author} {\bibfnamefont {L.}~\bibnamefont
  {Inno}}, \bibinfo {author} {\bibfnamefont {A.}~\bibnamefont {Rotundi}}, \
  and\ \bibinfo {author} {\bibfnamefont {A.}~\bibnamefont {Piccialli}},\ }\href
  {\doibase 10.1038/s41550-020-01258-z} {\bibfield  {journal} {\bibinfo
  {journal} {Nature Astronomy}\ }\textbf {\bibinfo {volume} {4}},\ \bibinfo
  {pages} {1114} (\bibinfo {year} {2020})}\BibitemShut {NoStop}%
\bibitem [{\citenamefont {{Leboulleux}}\ \emph {et~al.}(2021)\citenamefont
  {{Leboulleux}}, \citenamefont {{Cantalloube}}, \citenamefont {{Choquet}},
  \citenamefont {{Huby}},\ and\ \citenamefont {{Singh}}}]{Leboulleux2021}%
  \BibitemOpen
  \bibfield  {author} {\bibinfo {author} {\bibfnamefont {L.}~\bibnamefont
  {{Leboulleux}}}, \bibinfo {author} {\bibfnamefont {F.}~\bibnamefont
  {{Cantalloube}}}, \bibinfo {author} {\bibfnamefont {E.}~\bibnamefont
  {{Choquet}}}, \bibinfo {author} {\bibfnamefont {E.}~\bibnamefont {{Huby}}}, \
  and\ \bibinfo {author} {\bibfnamefont {G.}~\bibnamefont {{Singh}}},\ }in\
  \href@noop {} {\emph {\bibinfo {booktitle} {SF2A-2021: Proceedings of the
  Annual meeting of the French Society of Astronomy and Astrophysics. Eds.: A.
  Siebert}}},\ \bibinfo {editor} {edited by\ \bibinfo {editor} {\bibfnamefont
  {A.}~\bibnamefont {{Siebert}}}, \bibinfo {editor} {\bibfnamefont
  {K.}~\bibnamefont {{Bailli{\'e}}}}, \bibinfo {editor} {\bibfnamefont
  {E.}~\bibnamefont {{Lagadec}}}, \bibinfo {editor} {\bibfnamefont
  {N.}~\bibnamefont {{Lagarde}}}, \bibinfo {editor} {\bibfnamefont
  {J.}~\bibnamefont {{Malzac}}}, \bibinfo {editor} {\bibfnamefont {J.~B.}\
  \bibnamefont {{Marquette}}}, \bibinfo {editor} {\bibfnamefont
  {M.}~\bibnamefont {{N'Diaye}}}, \bibinfo {editor} {\bibfnamefont
  {J.}~\bibnamefont {{Richard}}}, \ and\ \bibinfo {editor} {\bibfnamefont
  {O.}~\bibnamefont {{Venot}}}}\ (\bibinfo {year} {2021})\ pp.\ \bibinfo
  {pages} {252--265}\BibitemShut {NoStop}%
\bibitem [{OEC(2022)}]{OECD2022}%
  \BibitemOpen
  \href {\doibase 10.1787/483507d6-en} {\enquote {\bibinfo {title} {First
  lessons from government evaluations of {COVID}-19 responses: A synthesis},}\
  } (\bibinfo {year} {2022})\BibitemShut {NoStop}%
\bibitem [{\citenamefont {{The World Bank}}(2021)}]{WB2021}%
  \BibitemOpen
  \bibfield  {author} {\bibinfo {author} {\bibnamefont {{The World Bank}}},\
  }\href@noop {} {\enquote {\bibinfo {title} {Gdp growth (annual \%)},}\ }
  (\bibinfo {year} {2021}),\ \bibinfo {note} {data retrieved from World
  Development Indicators,
  \url{https://data.worldbank.org/indicator/NY.GDP.MKTP.KD.ZG}}\BibitemShut
  {NoStop}%
\end{thebibliography}%

\appendix

\section{Monthly Analysis}
\label{app:A}
\begin{figure*}
\begin{center}
\includegraphics[width=0.48\textwidth]{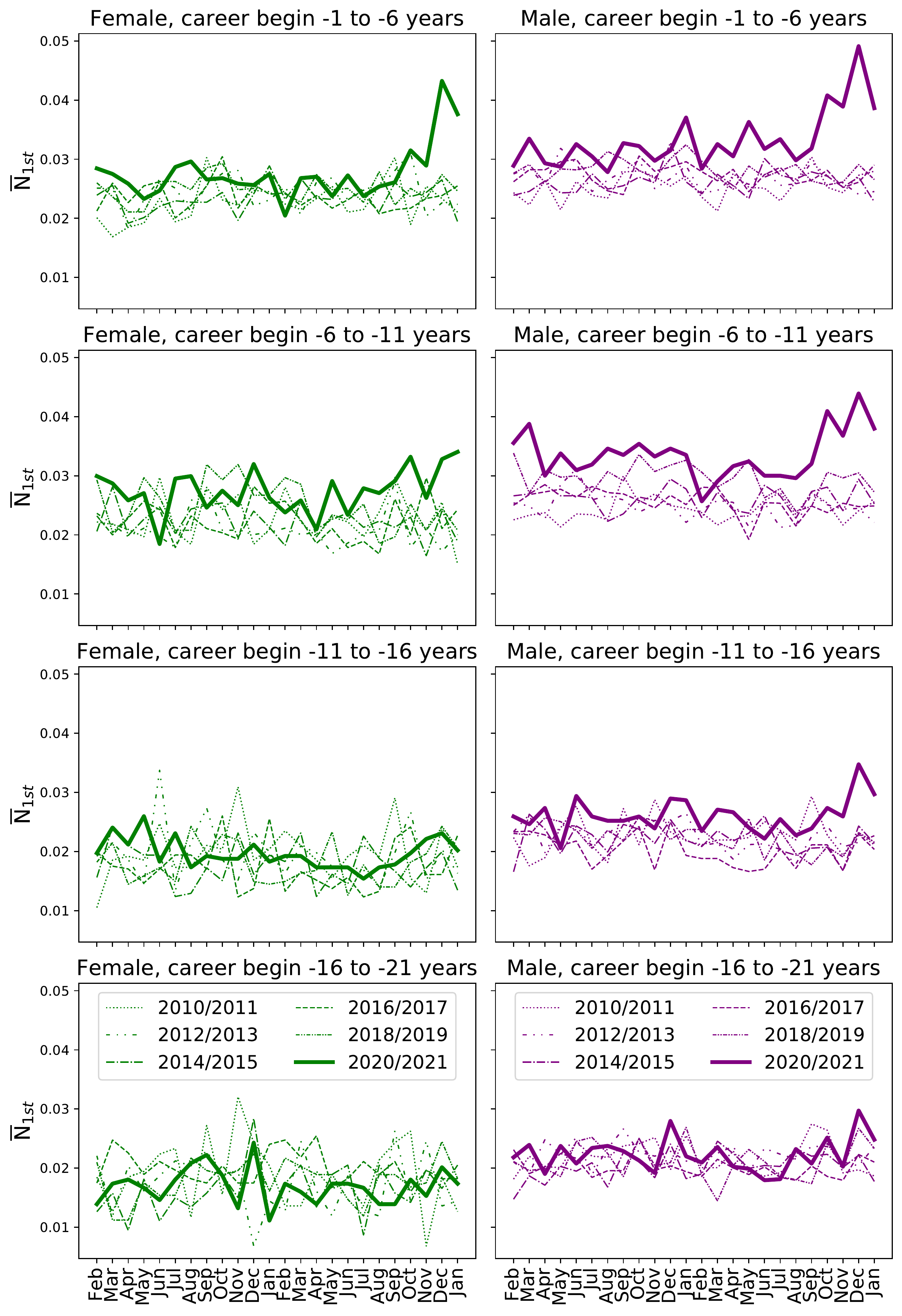}
\includegraphics[width=0.48\textwidth]{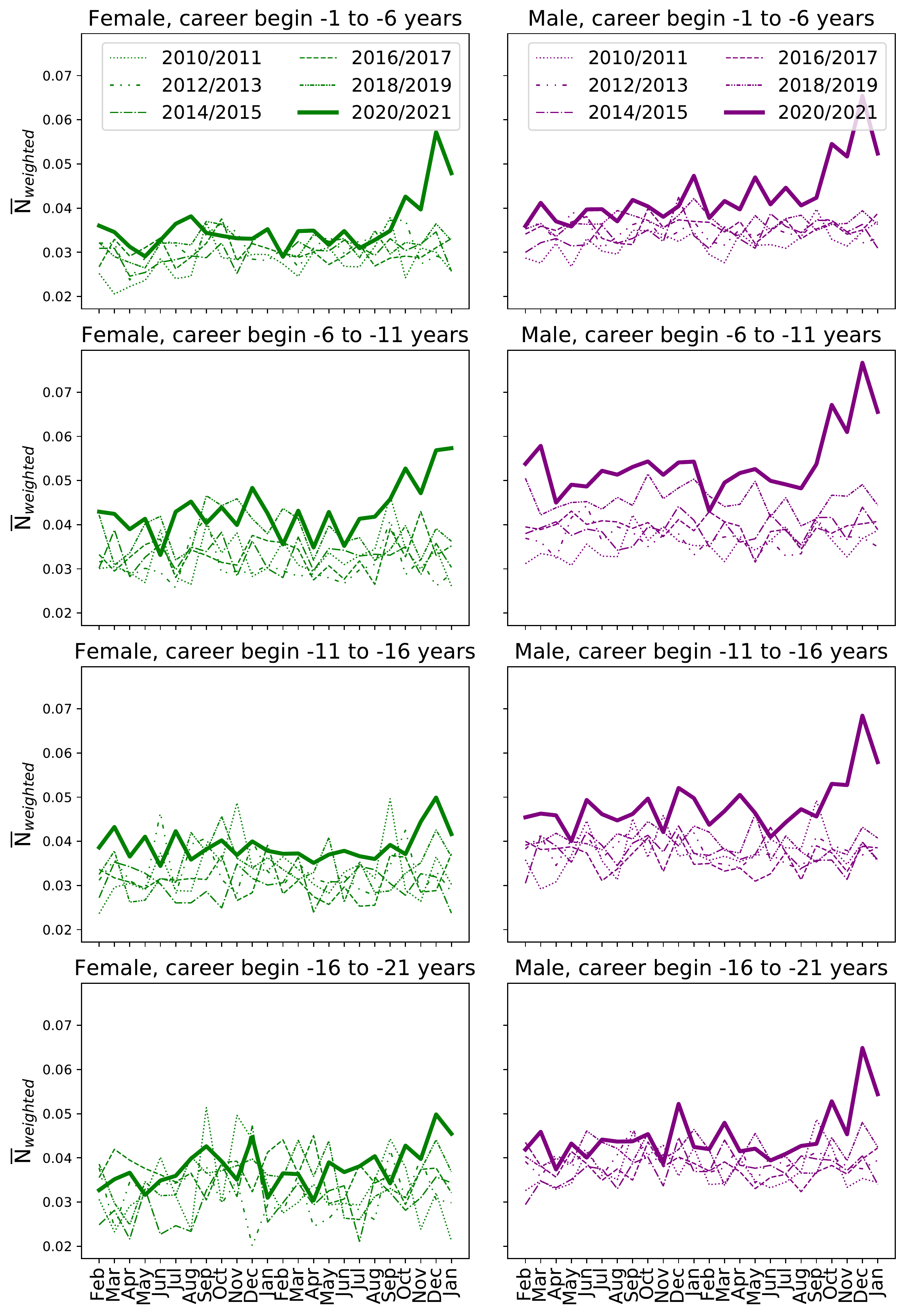}
\end{center}
\caption{\label{fig:monthly_nr} Average monthly productivity by gender and career stage. {\bf Left}: the average number of first author publications per author per month during a two-year interval. {\bf Right}: the average number of total publications per author weighted by author ranking. The thick solid line shows the productivity during COVID, thin lines correspond to time intervals pre-COVID. The y-scale is the same in all plots to ease the comparison between male (purple) and female (green) productivity.}
\end{figure*}
We conduct a monthly analysis of productivity during COVID (defined as the time interval February 2020--January 2022) and compare it to the productivity in previous two-year periods starting from 2010/2011.

To account for a possible dependence of productivity on career stage, we separate the results by career stage: For each time interval, we only consider publications from researchers that entered the field (had their first publication) within a certain period before the time interval under consideration. For example, for the time interval 2021/2022, we count all publications from researchers that entered between 2015--2019, while for the time interval 2019/2020 we count publications by authors who entered between 2013--2017. This ensures that we compare researchers at the same career stage.  

To correct for fluctuations in the number of active researchers in the field, we further normalize the results by dividing the number of publications by the number of researchers that entered the field during the chosen time window and that we also assume are still active during the time interval under consideration. We count authors as active if they published either during the two years before or during the time interval under consideration. We show result for 4 career stages, spanning a total of 20 year. We separate the results by gender and all publications versus refereed publications.

\begin{figure*}
\begin{center}
\includegraphics[width=0.48\textwidth]{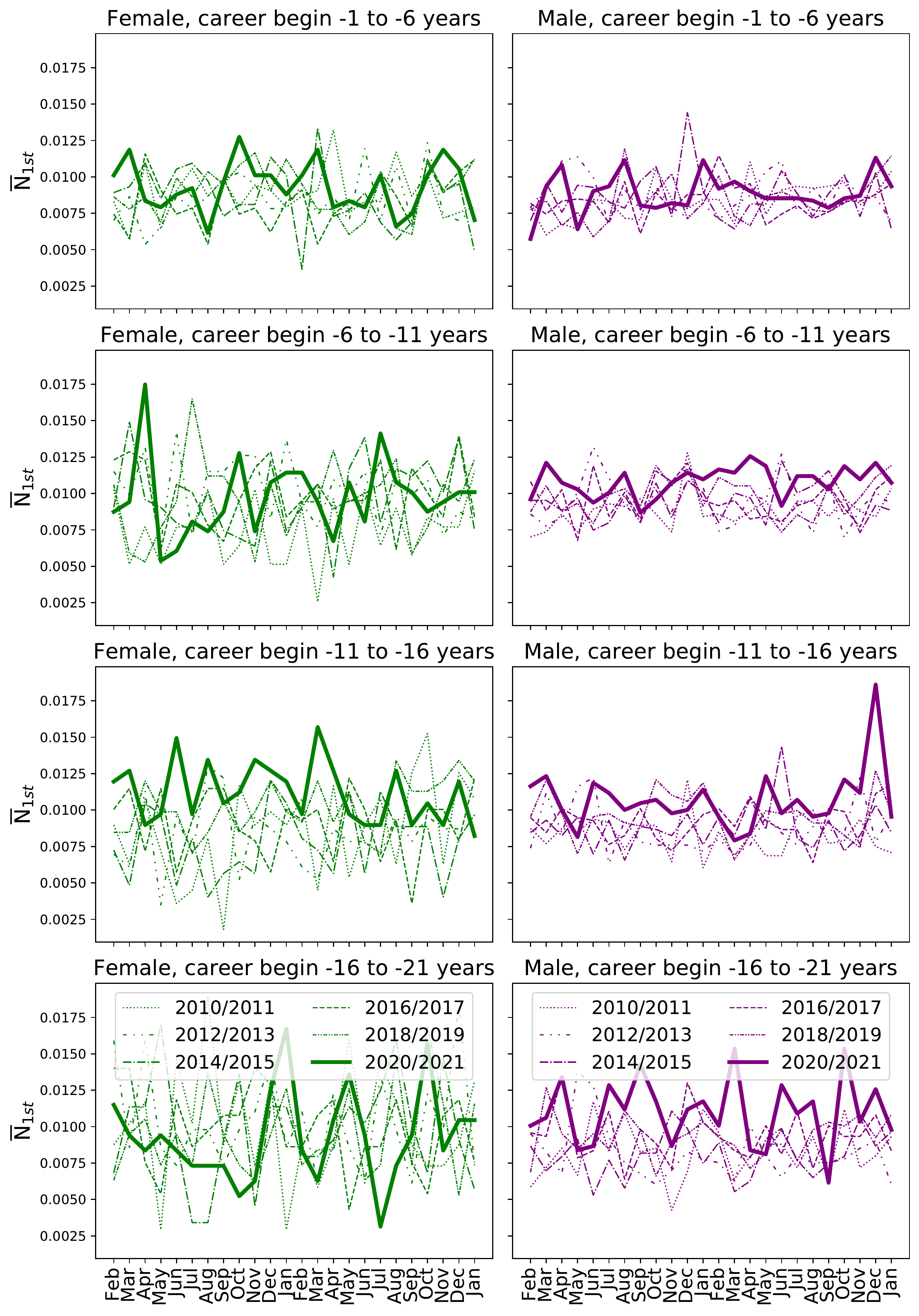}
\includegraphics[width=0.48\textwidth]{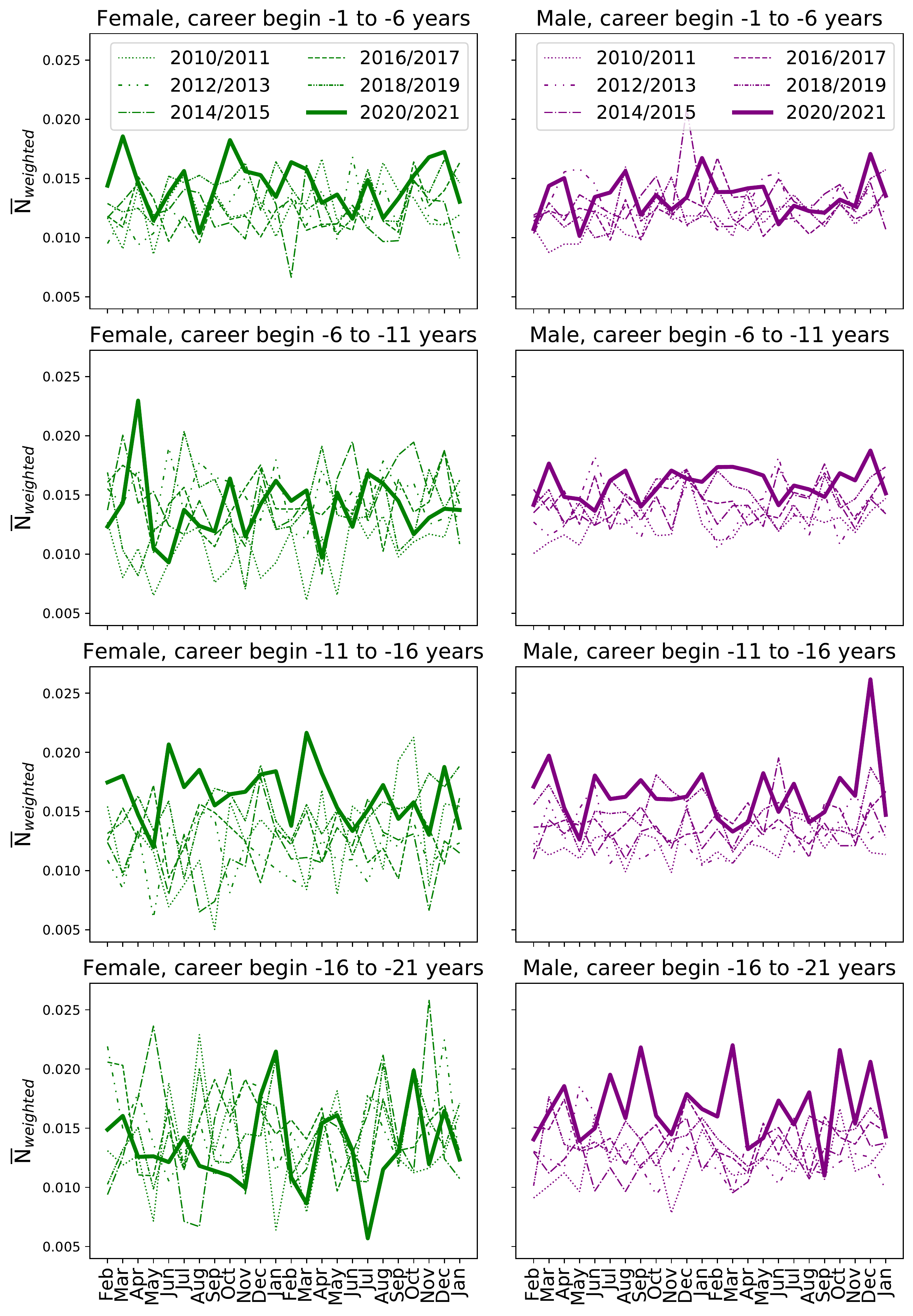}
\end{center}
\caption{\label{fig:monthly_r} Same as Fig.~\ref{fig:monthly_nr}, but restricted to refereed publications. The trends are weaker but still observable.}
\end{figure*}

The results of this analysis are shown in Fig.~\ref{fig:monthly_nr} (all publications) and Fig.~\ref{fig:monthly_r} (refereed publications only). Rows correspond to different career stages as indicated in the titles, columns separate results by gender and author ranking. The monthly analysis confirms the findings on the yearly data. We observe the productivity during COVID to be higher than during previous 2-year periods. The largest increase in productivity is seen for male astronomers at young and intermediate career stages and less so for women, indicating unequal shares in child and/or elderly care responsibilities or administrative work.
We find that the discrepancies are somewhat weaker in refereed publications, but still present. 

\end{document}